# An Assessment of the Radio Frequency Electromagnetic Field Exposure from A Massive MIMO 5G Testbed

Tian Hong Loh[1], Fabien Heliot[2], David Cheadle[1], Tom Fielder[1]

[1] Electromagnetic & Electrochemical Technologies Department, National Physical Laboratory, Teddington, United Kingdom,
tian.loh@npl.co.uk, david.cheadle@npl.co.uk, tom.fielder@npl.co.uk

[2] Institute for Communication Systems, University of Surrey, Guildford, United Kingdom, f.heliot@surrey.ac.uk

*Abstract*—Current radiofrequency electromagnetic field (RF-EMF) exposure limits have become a critical concern for fifth-generation (5G) mobile network deployment. Regulation is not harmonized and in certain countries and regions it goes beyond the guidelines set out by the International Commission on Non-Ionizing Radiation Protection (ICNIRP). Using a massive multiple-input-multiple-output (mMIMO) testbed with beamforming capabilities that is capable of mimicking realistic 5G base station (BS) performance, this paper presents an experimental and statistical assessment of its associated RF-EMF exposure within a real-world indoor environment. The mMIMO testbed has up to 128 channels with user-programmable software defined radio (SDR) capability. It could perform zero-forcing precoding after channel state information (CSI) acquisition for different beamforming scenarios with respect to the associated user terminal antenna setups and positions. With 64 active mMIMO transmit antennas, 8 beamforming scenarios have been considered for single-user (SU) and multi-user (MU) downlink communications at different locations. Using a calibrated triaxial isotropic field-probe, the received channel power heat map for each beamforming scenario was acquired and then converted into an RF-EMF heat map. The relevant RF-EMF statistics was evaluated based on the variations of beam profiles and number of users.

## I. INTRODUCTION

The fifth-generation (5G) of mobile networks, which promises high data rate, low latency and high reliability, is envisaged to be comprehensively rolled-out by 2020 [1]. The demand for high-speed communication for a range of diverse applications has driven strong global momentum in developing emerging 5G technologies to meet these needs. However, when it comes to the deployment of 5G, the radio-frequency electromagnetic field (RF-EMF) exposure limits have become a critical concern, especially in countries, regions and even specific cities where RF-EMF limits are significantly stricter than the International Commission on Non-Ionizing Radiation Protection (ICNIRP) guidelines [2], which are recommended by the World Health Organization (WHO). For example, in Italy and Poland, a different regulation is put in place where the current RF-EMF exposure limits are 6 V/m and 7 V/m, respectively, which is much stricter than the ICNIRP guidelines at 41 V/m [3].

Indeed, different countries/regions consider different precautionary principles and approaches to define the different radiation limitations per frequency band, per operator and per technology. The design of mobile networks based on RF-EMF exposure limits that are more restrictive than WHO recommendations, results in less flexibility when it comes to network deployment. Such more stringent exposure limits have already had an impact on 4G network rollout [3] and it is envisaged to be worse for 5G network deployment [4], [5].

The conventional measurements of RF-EMF exposure from third-generation (3G) and fourth-generation (4G) base stations (BSs) at the exclusion zone (a compliance boundary around the BS with no access to general public), which are based on the assumption that the theoretical maximum power is transmitted in each possible direction for a defined time-period, are becoming obsolete [6]. This is due to the complex new technology employed at 5G base stations such as beamforming massive multiple-input-multiple-output (mMIMO), which allows energy to be focused with sharp high gain beams in the direction of a specific mobile user resulting in non-realistically large exclusion zone areas. This makes it problematic for operators to deploy 5G Massive MIMO BSs on sites with pre-existing 3G and 4G BSs.

Regulators, operators and 5G equipment suppliers all require reliable and agreed assessment of RF-EMF exposure levels to support consistent and effective 5G regulation and network design. Scientific arguments and effective RF-EMF measurements on a massive MIMO system will be needed to support this vision. It is envisaged that suitable statistical approaches should form the base of RF-EMF exposure assessment for 5G new radio (NR) system employing mMIMO as beamforming in order to assure that high power user service beams are only transmitted on a need-to basis [4]–[6]. This paper presents an experimental assessment of RF-EMF exposure from a mMIMO 5G beamforming testbed within a real-world indoor environment that mimics the performance of a realistic 5G BS. The focus is to provide insights on how the RF-EMF is affected by the fluctuation of the environment, number and position of the different users and to evaluate the relevant statistics over the results obtained from measurement campaigns. This paper is organized as follows: Section II describes the experimental setups, Section III presents some measurement results and explains how these results have been used to generate a statistical model, and finally, conclusions are drawn in Section IV.

## II. EXPERIMENTAL SETUP

All the measurements were performed inside an indoor environment within a large meeting room, located at the basement of the 5G Innovation Centre (5GIC) at the University of Surrey [7]. The room is 15 m long, 7.5 m wide, and 3 m high. As depicted in Fig. 1, the typical furniture, such as chairs, and desk within the room were placed aside during the measurements. The room is surrounded by glass, brick and plasterboard walls. The floor defines the reference height $Z_0 = 0$ m; it is made of concrete and carpeted. The concrete suspended ceiling was equipped with some hanging lighting and projector equipment. During the measurement, all the measurement instruments were located inside the room. The following provides further details of the measurement instruments.

### A. mMIMO testbed and RF-EMF measurement system

The mMIMO testbed can perform phase-coherent and time synchronized MIMO baseband processing with user-programmable, reconfigurable and real-time signal processing field-programmable gate arrays (FPGAs)-based software defined radio (SDR) capabilities. The testbed consists of: 1) A BEE7 synchronization and trigger generator; 2) A MegaBEE transceiver modules (each module contents four input/output RF ports and could support up to 4 channels of IQ); 3) A White Rabbit time distribution system; 4) A transmit antenna array with 128 ($16 \times 8$) patch antenna elements; 5) Various receive antenna arrays with 4 dipole antenna element per user equipment (UE).

The synchronization of the mMIMO testbed is controlled by the BEE7 synchronization and a trigger generator. The signal generation and analysis are all implemented using the MegaBEE transceivers. The clocking network that achieves sub-nano second time synchronization between channels is derived from the White Rabbit time distribution system, which synchronizes a reference clock to each of the MegaBEE transceiver modules over an optical fibre link using SFP+ (Small Form-factor Pluggable Plus) network adaptors. Note that optical fibers are employed for both data transport and the clocking network.

The testbed provides flexible evaluation of various modulation schemes, new communication algorithms and protocols as well as enabling evaluation of the relevant over-the-air (OTA) link performance. For downlink communications, up to 128 channels could be used simultaneously at the transmitting end by using all the 32 transceiver modules whereas, at the receiving end, up to 32 channels can be used, i.e. up to 8 UE with 4 antenna elements each.

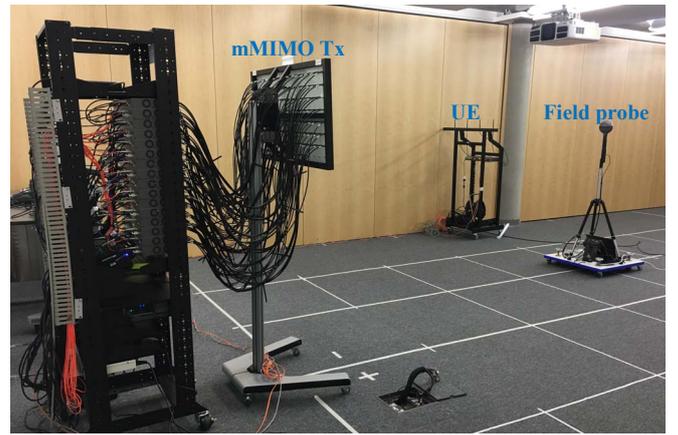

Fig. 1. RF-EMF assessment from a mMIMO testbed in an indoor environment.

The RF-EMF measurement system, in which the channel power could be acquired, consists of a handheld Keysight FieldFox N9917B portable spectrum analyser [8], and a an AGOS SDIA-6000 triaxial isotropic field probe [9] (allocated on a tripod as depicted in Fig. 1). To achieve traceability, this acquisition system has been calibrated at the Power Flux Density Laboratory in UK National Physical Laboratory (NPL) against known E-field [10].

### B. Instrument setup and test scenarios

The mMIMO testbed was programmed to perform downlink zero-forcing (ZF) precoding after channel state information (CSI) acquisition for different beamforming scenarios with respect to the associated user terminal antenna setups and positions. The RF-EMF measurement system was located on a trolley during the measurement (see Fig. 1).

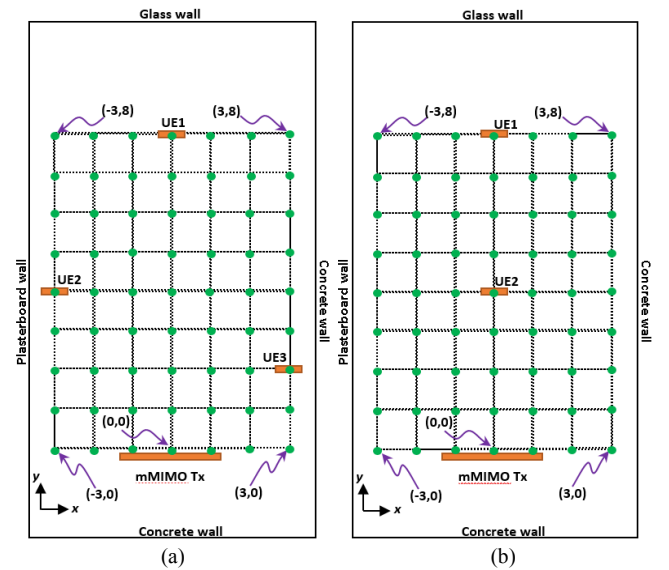

Fig. 2. Test scenarios for 64 active mMIMO Tx at (0 m,0 m): (a) 2 active UEs at (0 m, 8 m) and (0 m, 4 m); (b) 3 active UEs at (0 m, 8 m), (-3 m, 4 m) and (3 m, 2 m).

The mMIMO testbed was configured to operate with 64 active transmitting antennas. Each UE was operated with four vertically polarized dipole receiving antenna. The following beamforming scenarios for single-user (SU) and multi-user (MU) downlink communications at different locations have been considered: 1) SU at (0 m, 8 m); (2) SU at (-3 m, 4 m); 3) SU at (3 m, 2 m); 4) MU at (-3 m, 4 m) and (3 m, 2 m); 5) MU at (0 m, 8 m) and (0 m, 4 m); 6) MU at (0 m, 8 m) and (-3 m, 4 m); 7) MU at (0 m, 8 m) and (3 m, 2 m); 8) MU at (0 m, 8 m), (-3 m, 4 m) and (3 m, 2 m).

Figure 2 illustrates the two MU setup examples (i.e. Scenarios 5) and 8)), where the received channel power heat map was acquired by ensuring a consistent physical probe orientation throughout all the measured grid points (shown as green dots in the Fig. 2). The acquired channel power results for each beamforming scenario have then been converted into a corresponding RF-EMF heat map.

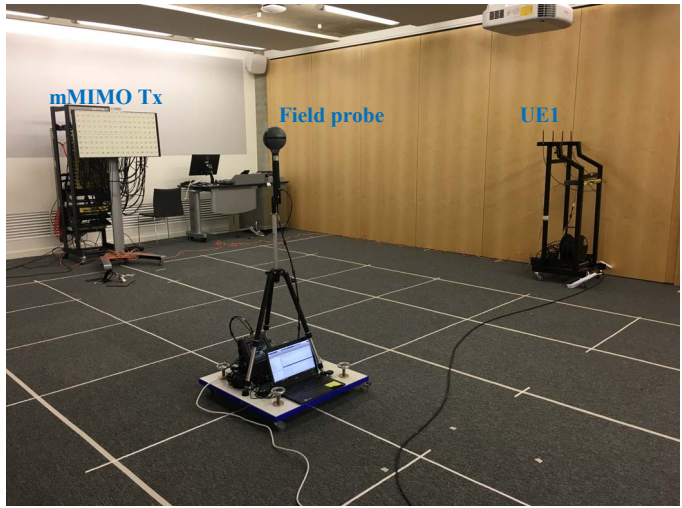

(a)

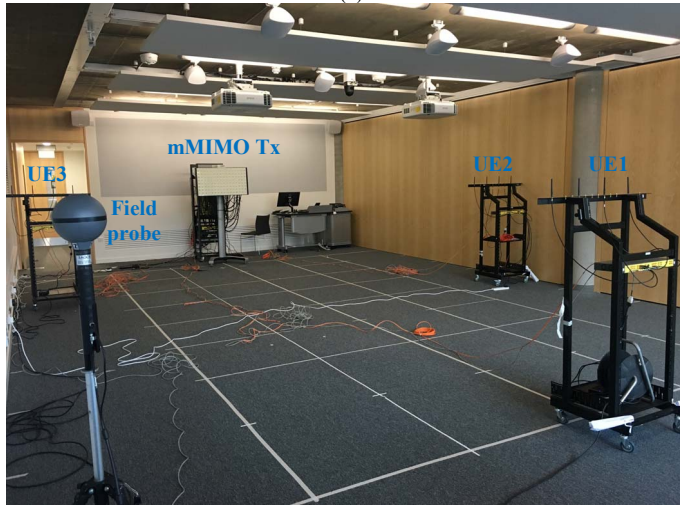

(b)

Fig. 3. Photos of the experimental setup for: (a) 1 active UE at (-3 m, 4 m); (b) 3 active UEs at (0 m, 8 m), (-3 m, 4 m) and (3 m, 2 m).

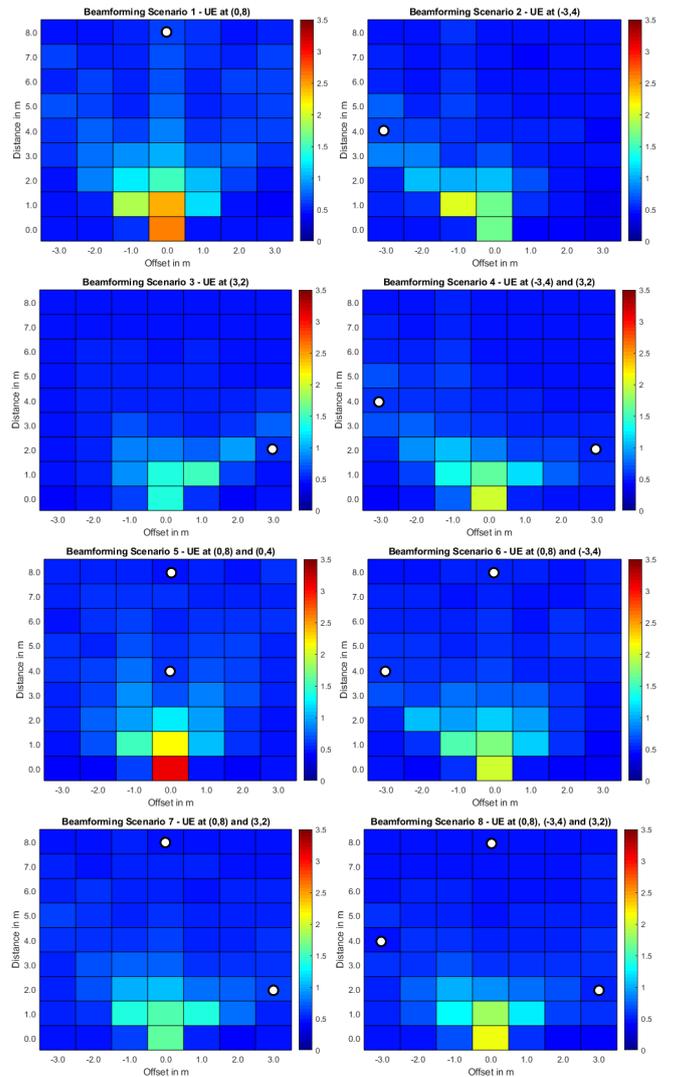

Fig. 4. Calibrated RF-EMF heat map [colour map value shown in V/m] for each different beamforming scenarios mentioned in Section II. Note that the white cicle shown in the plots depicts the UE locations.

Figure 3 shows photos of some relevant setups. The measurements were conducted at 2.63 GHz with a 40 MHz instantaneous data bandwidth per channel. All the baseband processing and algorithmic evaluation are performed in Matlab. A fixed length of 65536 samples (i.e. $2^{16}$) was set to transmit at 61.44 Msps (mega-samples per second) from the multiple antenna transmitter. Orthogonal frequency-division multiplexing (OFDM) frames were generated with subcarrier spacing of 15 kHz and 64 quadrature amplitude modulation (64QAM) modulation in time division duplex (TDD) mode.

For the 8 beamforming scenarios, the same measurement procedure was followed: 1) first a pilot frame is sent from the BS to each UE, which acquires and feedback its CSI via the SFP+ cable; 2) the BS then estimates the CSI and uses it to generate ZF precoded data frame; 3) finally the BS transmits theses ZF precoded data frames in a continuous manner while the probe is moved to the 56 different grid positions of Fig. 2 to measure the EMF. In order to ensure that the ZF precoding reasonably well, the uncoded bit-error-rate (BER) when using a 64-QAM is calculated at each UE based on the first received data frame.

## III. EXPERIMENTAL RESULTS

Table I presents the BER values at each UE for each beamforming scenario; the results show that when using ZF precoding, the more the users, the lower the quality of the beamforming. But still in any case uncoded BER of around $10^{-3}$ can be achieved, which tend to confirm that the BS beamforms in the expected direction.

TABLE I.  BER OF EACH UE FOR EACH BEAMFORMING TEST SCENARIO LISTED IN SECTION II *B*

| Scenario | UE1 | UE2 | UE3 |
|---|---|---|---|
| 1) | $1.35 \times 10^{-6}$ | N/A | N/A |
| 2) | $4.61 \times 10^{-6}$ | N/A | N/A |
| 3) | $5.12 \times 10^{-6}$ | N/A | N/A |
| 4) | $2.35 \times 10^{-6}$ | $4.82 \times 10^{-5}$ | N/A |
| 5) | $8.38 \times 10^{-5}$ | $9.22 \times 10^{-5}$ | N/A |
| 6) | $3.68 \times 10^{-5}$ | $4.31 \times 10^{-5}$ | N/A |
| 7) | $4.37 \times 10^{-5}$ | $3.99 \times 10^{-5}$ | N/A |
| 8) | $8.04 \times 10^{-4}$ | $2.68 \times 10^{-3}$ | $5.11 \times 10^{-4}$ |

Figure 4 presents the calibrated RF-EMF heat map in *x-y* coordinate in V/m for the eight different beamforming scenarios mentioned in Section II. From the heat map results, the envisaged 'beam profile' has been observed. By considering all the beamforming scenarios (i.e. varying beam profile and number of users), the average RF-EMF heat map is presented in Fig. 5. Both Figs. 4 and 5 illustrate that even though we know from Table I that the BS beamforms in the expected direction in each test scenario (since reasonably low BER can be achieved at each UE) the highest RF-EMF exposure from the 64 Tx mMIMO testbed is around the coordinate (0 m, 0 m) in most cases, where the mMIMO was located.

To assess the statistical insight, a cut of the calibrated RF-EMF heat map for the coordinate – '*x* = 0 m and *y* varying between 0 m and 8 m 'was chosen. The average and the relevant RF-EMF for each beamforming scenarios are presented in Figure 6. The results illustrate that, by varying the beam profile and number of users, the maximum RF-EMF exposure from the 64 Tx mMIMO testbed at (0 m, 0 m) varies between 1.37 V/m and 3.09 V/m. A general trend on the exposure from mMIMO testbed decreases according to the inverse square law, i.e. 1/(distance) is also observed even when the main beam is not pointing at direction along *y*-axis from (0 m, 0 m) (i.e. Scenario 4).

Further work envisages to be carried out are to investigate the effect on RF-EMF statistics by also considering the variation of user data traffic, number of active Tx from mMIMO testbed as well as to consider the presence of obstacles, high population UE density, introduce interference sources, assessment in different test environments, etc.

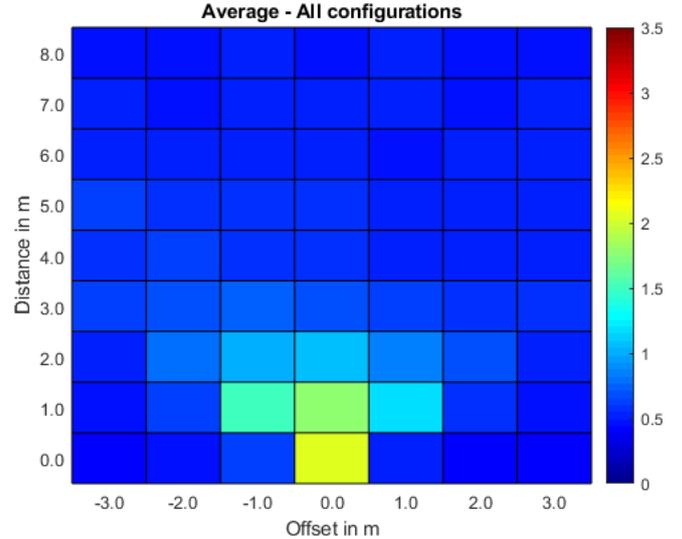

Fig. 5. Averaged RF-EMF heat map [colour map value shown in V/m] for all the eight different beamforming scenarios mentioned in Section II.

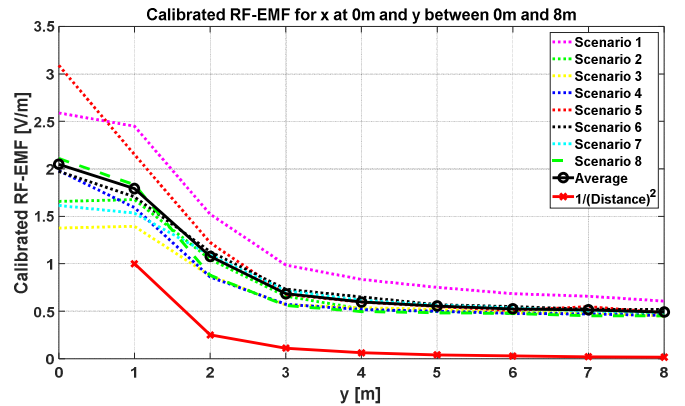

Fig. 6. RF-EMF at $x$ = 0 m and $y$ between 0 m to 8 m.

## IV. CONCLUSION

This paper has presented an experimental assessment of RF-EMF exposure at 2.63 GHz from a mMIMO testbed with ZF precoding SDR ability within a real-world indoor environment. With 64 active mMIMO transmit antennas, 8 beamforming scenarios have been considered for SU and MU downlink communications at different locations. The associated channel power heat map was acquired by using a triaxial field-probe, which was then converted to RF-EMF. It is envisaged that the work will lead to more informed debate and regulations, that balance performance and public safety. Sound evidence will also assist the regulators to decide whether to harmonise policies on deploying future mobile broadband and wireless technologies.

ACKNOWLEDGMENT

This work was supported by the EU project 5GRFEX entitled – 'Metrology for RF exposure from Massive MIMO 5G base station: Impact on 5G network deployment' (this project has received funding from the support for impact (SIP) programme co-financed by the Participating States and from the European Union's Horizon 2020 research and innovation programme), under European Association of National Metrology Institutes (EURAMET) Reference 18SIP02. The authors would like to thank Mr Lawrence Carslake for his assistance over the measurements.